\documentclass {elsarticle}
\usepackage [utf8] {inputenc}
\usepackage [T1] {fontenc}
\usepackage {url}

\usepackage{listings}
\usepackage{xcolor}

\definecolor{dkred}{rgb}{0.5,0.1,0}
\definecolor{dkblue}{rgb}{0,0.1,0.5}
\definecolor{ltblue}{rgb}{0,0.5,0.5}
\definecolor{dkgreen}{rgb}{0,0.6,0}
\definecolor{dk2green}{rgb}{0,0.2,0}
\definecolor{dkviolet}{rgb}{0.6,0,0.8}

\newcommand{\code}[1]{\texttt{#1}}

\usepackage {lstcoq}

\begin {document}

\begin {frontmatter}
\title {Formalization of context-free language theory}
\author{Marcus V. M. Ramos}	
\ead{mvmr@cin.ufpe.br}
\address{Centro de Informática, UFPE, Recife, Brazil} 

\author{Ruy J. G. B. de Queiroz}
\ead{ruy@cin.ufpe.br}
\address{Centro de Informática, UFPE, Recife, Brazil} 

\author{Nelma Moreira}
\ead{nam@dcc.fc.up.pt}
\address{Departamento de Ciência de Computadores, Faculdade de Ciências, Universidade do Porto, Porto, Portugal} 

\author{José Carlos Bacelar Almeida}
\ead {jba@di.uminho.pt}
\address{Departamento de Informática, Universidade do Minho, Braga, Portugal} 

\begin {abstract}
Context-free language theory is a subject of high importance in computer language processing technology as well as in formal language theory. This paper presents a formalization, using the Coq proof assistant, of fundamental results related to context-free grammars and languages. These include closure properties (union, concatenation and Kleene star), grammar simplification (elimination of useless symbols inaccessible symbols, empty rules and unit rules) and the existence of a Chomsky Normal Form for context-free grammars.
\end {abstract}

\end {frontmatter}

\section {Introduction}
\label {sec-intro}
The formalization of context-free language theory is a key to the certification of compilers and programs, as well as to the development of new languages and tools for certified programming. 

The objective of this work is to formalize a substantial part of context-free language theory in the Coq proof assistant, making it possible to reason about it in a fully checked environment, with all the related advantages. Initially, however, the focus has been restricted to context-free grammars and associated results. Pushdown automata and their relation to context-free grammars are not considered.

In order to follow this paper, the reader is required to have basic knowledge of Coq and of context-free language theory. For the beginner, the recommended starting point for Coq is the book by Bertot and Castéran \cite {bertot-2004}. Background on context-free language theory can be found in \cite {sudkamp-2006} or \cite {ramos-2009}, among others.

The general idea of formalizing context-free language theory in the Coq proof assistant is discussed in Section \ref {sec-basic}. A library that contains fundamental results on context-grammars, and supports the whole formalization, is briefly presented in Section \ref {sec-library}. In Section \ref {sec-methodology} we give an overview of the work, by explaining the common approach adopted in its different parts. Specific results related to the formalization of closure properties of context-free languages, grammar simplification and Chomsky Normal Form are presented, respectively, in Sections \ref {sec-clopro}, \ref {sec-simplification} and \ref {sec-cnf}. Section \ref {sec-related} discusses related work by various other researchers and final conclusions are presented in Section \ref {sec-conclusions}.

As far as the authors are aware of, this is the first comprehensive formalization of important results of context-free language theory in the Coq proof assistant. Previous publications of the authors presented the formalization of closure properties for context-free grammars (in an earlier version) \cite {ramos-2014} and of simplification for context-free grammars \cite {ramos-2015}. All the definitions and proof scripts discussed in this paper were written in plain Coq and are available for download at: \\
\url {https://github.com/mvmramos/chomsky} \\

\section {Basic Definitions}
\label {sec-basic}

In this section we present how the main concepts and objects of context-free language theory were defined in our formalization in Coq. They are used throughout the work.

\subsection {Grammars}
Context-free grammars were represented in Coq very closely to the usual algebraic definition $G=(V,\Sigma,P,S)$, where $V$ is the vocabulary of $G$ (it includes all non-terminal and terminal symbols), $\Sigma$ is the set of terminal symbols (used in the construction of the sentences of the language generated by the grammar), $N=V\setminus\Sigma$ is the set of non-terminal symbols (representing different sentence abstractions), $P$ is the set of rules and $S \in N$ is the start symbol (also called initial or root symbol). Rules have the form $\alpha \rightarrow \beta$, with $\alpha \in N$ and $\beta \in V^*$.

Basic definitions in Coq are presented below. The $N$ and $\Sigma$ sets are represented separately from $G$ (respectively, by types \texttt {non\_terminal} and \texttt {terminal}). Notations \texttt {sf} (sentential form) and \texttt {sentence} represent lists, possibly empty, of respectively terminal and non-terminal symbols and terminal only symbols.

\begin{coq}
Variables non_terminal terminal: Type.
Notation sf := (list (non_terminal + terminal)).
Notation sentence := (list terminal).
Notation nlist:= (list non_terminal).
\end{coq}

The record representation \texttt {cfg} has been used for $G$. The definition states that \texttt {cfg} is a new type and contains three components. The first is the \texttt {start\_symbol} of the grammar (a non-terminal symbol) and the second is \texttt {rules}, that represents the rules of the grammar. Rules are propositions (represented in Coq by \texttt {Prop}) that take as arguments a non-terminal symbol and a (possibly empty) list of non-terminal and terminal symbols (corresponding, respectively, to the left and right-hand side of a rule). 

The predicate \texttt {rules\_finite\_def} assures that the set of rules of the grammar is finite by proving that the length of right-hand side of every rule is equal or less than a given value, and also that both left and right-hand side of the rules are built from finite sets of, respectively, non-terminal and terminal symbols (represented here by lists).

\begin{coq}
Definition rules_finite_def 
   (non_terminal terminal : Type)
   (ss: non_terminal) 
   (rules: non_terminal -> sf -> Prop) 
   (n: nat) 
   (ntl: list non_terminal)
   (tl: list terminal) :=
In ss ntl /\
(forall left: non_terminal,
 forall right: list (non_terminal + terminal),
 rules left right ->
 length right <= n /\
 In left ntl /\
 (forall s : non_terminal, In (inl s) right -> In s ntl) /\
 (forall s : terminal, In (inr s) right -> In s tl)).

Record cfg (non_terminal terminal : Type): Type:= {
start_symbol: non_terminal;
rules: non_terminal -> sf -> Prop;
rules_finite: 
   exists n: nat,
   exists ntl: nlist,
   exists tl: tlist,
   rules_finite_def start_symbol rules n ntl tl }.
\end{coq}

The decision of representing rules as propositions has the consequence that it will not allow for direct extraction of executable code from the formalization. It would surely be desirable, however, to be able to obtain certified algorithms for the operations described in this article. The alternative, in this case, would be to represent rules as a member of type \texttt {list (non\_terminal * sf)} instead. This, however, would have changed the declarative approach of the present work into the algorithmic approach, by creating functions that generate new grammars with the desired properties. The purely logical approach was considered more appealing, since its maps directly from the textbooks, and thus was selected as the choice for the present formalization. In any case, it does not affect the objectives listed in Section \ref {sec-intro} and can be adapted in the future in order to allow for code extraction, although this should demand a considerable effort in the creation and proof of program-related scripts.

The example below represents the grammar $$G=(\{S',A,B,a,b\},\{a,b\},\{S' \rightarrow aS', S' \rightarrow b\},S')$$ that generates language $a^*b$:

\begin{coq}
Inductive nt1: Type:= | S' | A | B.
Inductive t1: Type:= | a | b.
Inductive rs1: nt1 -> list (nt1 + t1) -> Prop:=
  r1: rs1 S' [inr a; inl S']
| r2: rs1 S' [inr b].

Definition g1: cfg nt1 t1:= {|
start_symbol:= S'; 
rules:= rs1;
rules_finite:= rs1_finite |}.
\end{coq}

The term \texttt {rs1\_finite} (the proof that the set of rules of \texttt {g1} is finite) is not presented here, but can be easily constructed and is available from the link provided in Section \ref {sec-intro}. 

\subsection {Derivations}
Another fundamental concept used in this formalization is the idea of \emph {derivation}: a grammar \texttt{g} \emph {derives} a string \texttt {s2} from a string \texttt {s1} if there exists a series of rules in \texttt {g} that, when applied to \texttt {s1}, eventually result in \texttt {s2}. A direct derivation (i.e, the application of a single rule) is represented by $s_1 \Rightarrow s_2$, and the reflexive and transitive closure of this relation (i.e, the application of zero or more rules) is represented by $s_1 \Rightarrow^* s_2$. An inductive predicate definition of this concept in Coq (\texttt {derives}) uses two constructors:

\begin{coq}
Inductive derives 
   (non_terminal terminal : Type) 
   (g : cfg non_terminal terminal)
   : sf -> sf -> Prop :=
   | derives_refl : 
      forall s : sf, 
      derives g s s
   | derives_step : 
      forall (s1 s2 s3 : sf)
      forall (left : non_terminal)
      forall (right : sf),
      derives g s1 (s2 ++ inl left :: s3) ->
      rules g left right -> derives g s1 (s2 ++ right ++ s3)
\end{coq}

The constructors of this definition (\texttt {derives\_refl} and \texttt {derives\_step}) are the axioms of our theory. Constructor \texttt {derives\_refl} asserts that every sentential form \texttt {s} can be derived from \texttt {s} itself. Constructor \texttt {derives\_step} states that if a sentential form that contains the left-hand side of a rule is derived by a grammar, then the grammar derives the sentential form with the left-hand side replaced by the right-hand side of the same rule. This case corresponds to the application of a rule in a direct derivation step.

A grammar \texttt {generates} a string if this string can be derived from its start symbol. Finally, a grammar \texttt {produces} a sentence if it can be generated from its start symbol.

\begin{coq}
Definition generates (g: cfg) (s: sf): Prop:=
derives g [inl (start_symbol g)] s.

Definition produces (g: cfg) (s: sentence): Prop:=
generates g (map terminal_lift s).
\end{coq} 

Function \texttt {terminal\_lift} converts a terminal symbol into an ordered pair of type \texttt {(non\_terminal + terminal)}. With these definitions, it has been possible to prove various lemmas about grammars and derivations, and also operations on grammars, all of which were useful when proving the main theorems of this article. 

As an example, the lemma that states that $G$ produces the string $aab$ (that is, that $aab \in L(G)$) is represented as:

\begin{coq}
Lemma g1_produces_aab:
produces g1 [a; a; b].
\end{coq} 

The proof of this lemma can be easily constructed and relates directly to the derivations in $S \Rightarrow aS \Rightarrow aaS \Rightarrow aab$, however in reverse order because of the way that \texttt {derives} is defined.

\subsection {Languages}
\label {sub-languages}

A language is a set of strings over a given alphabet. It is also useful to define the language generated by a grammar as the set of terminal strings generated by the grammar through derivations: $$L(G) = \{w\,|\,S \Rightarrow^*_g w\}$$

From the formalization point of view, we have defined language as a function that is parametrized over a certain type (representing the set of terminal symbols), takes a string built from elements of this type and returns a proposition asserting that the string belongs to the language:

\begin{coq}
Definition lang (terminal: Type):= sentence -> Prop.
\end{coq} 

The language generated by a grammar is then a function whose return value is the predicate \texttt {produces} presented earlier:

\begin{coq}
Definition lang_of_g (g: cfg): lang :=
fun w: sentence => produces g w.
\end{coq} 

Finally, we are able to define the equality of two languages, and also the fact that a certain language is a context-free language (which, in this case, means that there exists a context-free grammar that generates this language):

\begin{coq}
Definition lang_eq (l k: lang) := 
forall w, l w <-> k w.

Infix "==" := lang_eq (at level 80). 

Definition cfl (terminal: Type) (l: lang terminal): Prop:=
exists non_terminal: Type, 
exists g: cfg non_terminal terminal, 
l == lang_of_g g.
\end{coq} 

The symbol \texttt {==} is a notation for \texttt {lang\_eq}. 

Two grammars $g_1$ (with start symbol $S_1$) and $g_2$  (with start symbol $S_2$) are \emph {equivalent} (denoted $g_1 \equiv g_2$) if they generate the same language, that is, $\forall s, (S_1 \Rightarrow^*_{g_1} s) \leftrightarrow (S_2 \Rightarrow^*_{g_2} s)$. This is represented in our formalization in Coq by the predicate \texttt {g\_equiv}:

\begin{coq}
Definition g_equiv 
(non_terminal1 non_terminal2 terminal : Type)
(g1: cfg non_terminal1 terminal) 
(g2: cfg non_terminal2 terminal): Prop:=
forall s: sentence,
produces g1 s <-> produces g2 s.
\end{coq}

\section {Generic CFG Library}
\label {sec-library}

The definitions presented in the previous section allowed the construction of a generic library of fundamental lemmas on context-free grammars. This library was later used in the formalization of the specific results discussed next. It includes, among others, the formalization of the following statements and the corresponding proofs:

\begin {itemize}
\item $\forall g, s_1, s_2, s_3,\, (s_1 \Rightarrow^*_g s_2) \rightarrow (s_2 \Rightarrow^*_g s_3) \rightarrow (s_1 \Rightarrow^*_g s_3)$
\item $\forall g, s_1, s_2, s, s',\, (s_1 \Rightarrow^*_g s_2) \rightarrow (s \cdot s_1 \cdot s' \Rightarrow^*_g s \cdot s_2 \cdot s')$
\item $\forall g, s_1, s_2, s_3, s_4,\, (s_1 \Rightarrow^*_g s_2) \rightarrow (s_3 \Rightarrow^*_g s_4) \rightarrow (s_1 \cdot s_3 \Rightarrow^*_g s_2 \cdot s_4)$
\item $\forall g, s_1, s_2, s_3,\, \\
(s_1 \cdot s_2 \Rightarrow^*_g s_3) \rightarrow \exists s_1', s_2'\,|\,(s_3 = s_1' \cdot s_2') \land (s_1 \Rightarrow^*_g s_1') \land (s_2 \Rightarrow^*_g s_2')$
\item $\forall g, s_1, s_2, n, w,\, (s_1 \cdot n \cdot s_2 \Rightarrow^*_g w) \rightarrow \exists w'\,|\,(n \Rightarrow^*_g w')$
\item $\forall g, n, w,\, (n \Rightarrow^*_g w) \rightarrow (n \rightarrow_g w) \lor (\exists right\,|\,n \rightarrow_g right \land right \Rightarrow^*_g w)$
\item $\forall g_1, g_2, g_3,\, (g_1\equiv g_2) \land (g_2\equiv g_3) \rightarrow (g_1\equiv g_3)$
\end {itemize}

\noindent where $s, s', s_1, s_1', s_2', s_2, s_3, s_4$ and $right$ are sentential forms, $n$ is a non-terminal symbol, $w$ is a sentence and $g_1, g_2$ and $g_3$ are context-free grammars.

Also, additional notions of derivations were defined in order to simplify the proofs. This is the case, for example, where a reduction (the opposite of a derivation) or an induction on the number of derivation steps are required. For these cases, we have defined $derives2$ (which reduces a sentential form according to a rule) and $derives6$ (which controls the number of rules applied in the derivation):

\begin{coq}
Inductive derives2 
   (non_terminal terminal : Type)
   (g : cfg non_terminal terminal)
   : sf -> sf -> Prop :=
   | derives2_refl : 
      forall s : sf, 
      derives2 g s s
   | derives2_step : 
      forall (s1 s2 s3 : sf)
      forall (left : non_terminal)
      forall (right : sf),
      derives2 g (s1 ++ right ++ s2) s3 ->
      rules g left right ->
      derives2 g (s1 ++ inl left :: s2) s3
	  
Inductive derives6 
   (non_terminal terminal : Type)
   (g : cfg non_terminal terminal)
   : nat -> sf -> sf -> Prop :=
   | derives6_0 : 
      forall s : sf, 
      derives6 g 0 s s
   | derives6_sum : 
      forall (left : non_terminal)
      forall (right : sf) 
      forall (i : nat)
      forall (s1 s2 s3 : sf),
      rules g left right ->
      derives6 g i (s1 ++ right ++ s2) s3 ->
      derives6 g (S i) (s1 ++ [inl left] ++ s2) s3
\end{coq}

For the first case, we proved $derives\;g\;s_1\;s_2 \leftrightarrow derives2\;g\;s_1\;s_2$. For the second case, we proved $derives\;g\;s_1\;s_2 \leftrightarrow \exists n, derives6\;g\;n\;s_1\;s_2$

\section {Methodology}
\label {sec-methodology}

This formalization is essentially about context-free grammar manipulation. That is, about the definition of a new grammar from a previous one (or two), such that it satisfies some very specific properties. This is exactly the case when we define new grammars that generate the union, concatenation, closure (Kleene star) of given input grammar(s). Also, when we create new grammars that exclude empty rules, unit rules, useless symbols and inaccessible symbols from the original ones. Finally, when we propose a new grammar based on some other grammar, that satisfies a specific normalization standard, the Chomsky Normal Form.

For all these cases, the following approach has been adopted:

\begin {enumerate}
\item Depending on the case, define a new type of non-terminal symbols; this will be important, for example, when we want to guarantee that the start symbol of the grammar does not appear in the right-hand side of any rule or when we have to construct new non-terminals from the existing ones;
\item Inductively define the rules of the new grammar, in a way that allows the construction of the proofs that the resulting grammar has the required properties; these new rules will likely make use of the new non-terminal symbols described above;
\item Define the new grammar by using the new non-terminal symbols and the new rules; define the new start symbol (which might be a new symbol or an existing one) and build a proof of the finiteness of the set of rules for this new grammar;
\item State and prove all the lemmas and theorems that will assert that the newly defined grammar has the desired properties;
\item Consolidate the results within the same scope and finally with the previously obtained results.
\end {enumerate}

In the following sections, this approach will be explored with further detail for each main result achieved in this work.

\section {Closure Properties}
\label {sec-clopro}

After context-free grammars and derivations were defined, and the generic CFG library was built, the basic operations of concatenation, union and closure for context-free grammars were described in a rather straightforward way. These operations provide, as their name suggests, new context-free grammars that generate, respectively, the concatenation, the union and the closure of the language(s) generated by the input grammar(s). 

\subsection {Union}

Given two arbitrary context-free grammars $g_1$ and $g_2$, the following definitions are used to construct $g_3$ such that $L(g_3)=L(g_1) \cup L(g_2)$ (that is, the language generated by $g_3$ is the union of the languages generated by $g_1$ and $g_2$).

The first definition below (\texttt {g\_uni\_nt}) represents the type of the non-terminal symbols of the union grammar, created from the non-terminal symbols of the source grammars (respectively, \texttt {non\_terminal1} and \texttt {non\_terminal2}). Initially, the non-terminals of the source grammars are mapped to non-terminals of the union grammar. Second, there is the need to add a new and unique non-terminal symbol (\texttt {Start\_uni}), which will be the start symbol of the union grammar.

The functions \texttt {g\_uni\_sf\_lift1} and \texttt {g\_uni\_sf\_lift2} simply map sentential forms from, respectively, the first or the second grammar, and produce sentential forms of the union grammar. This will be useful when defining the rules of the union grammar.

\begin{coq}
Inductive g_uni_nt (non_terminal_1 non_terminal_2 : Type): Type:=
| Start_uni
| Transf1_uni_nt: non_terminal_1 -> g_uni_nt
| Transf2_uni_nt: non_terminal_2 -> g_uni_nt.

Notation sf1:= (list (non_terminal_1 + terminal)).
Notation sf2:= (list (non_terminal_2 + terminal)).
Notation sfu:= (list (g_uni_nt + terminal)).

Definition g_uni_sf_lift1 (c: non_terminal_1 + terminal)
: g_uni_nt + terminal:=
  match c with
  | inl nt => inl (Transf1_uni_nt nt)
  | inr t  => inr t
  end.

Definition g_uni_sf_lift2 (c: non_terminal_2 + terminal)
: g_uni_nt + terminal:=
  match c with
  | inl nt => inl (Transf2_uni_nt nt)
  | inr t  => inr t
  end.
\end{coq}

The rules of the union grammar are represented by the inductive definition \texttt {g\_uni\_rules}. Constructors \texttt {Start1\_uni} and \texttt {Start2\_uni} state that two new rules are added to the union grammar: respectively the rule that maps the new start symbol to the start symbol of the first grammar, and the rule that does the same for the second grammar. Then, constructors \texttt {Lift1\_uni} and \texttt {Lift2\_uni} simply map rules of first (resp. second) grammar into rules of the union grammar.

\begin{coq}
Inductive g_uni_rules 
(non_terminal_1 non_terminal_2 terminal : Type)
(g1: cfg non_terminal_1 terminal) 
(g2: cfg non_terminal_2 terminal)
: g_uni_nt -> sfu -> Prop :=
| Start1_uni: 
   g_uni_rules g1 g2 Start_uni [inl (Transf1_uni_nt (start_symbol g1))]
| Start2_uni: 
   g_uni_rules g1 g2 Start_uni [inl (Transf2_uni_nt (start_symbol g2))]
| Lift1_uni: 
   forall nt: non_terminal_1,
   forall s: sf1,
   rules g1 nt s ->
   g_uni_rules g1 g2 (Transf1_uni_nt nt) (map g_uni_sf_lift1 s)
| Lift2_uni: 
   forall nt: non_terminal_2,
   forall s: sf2,
   rules g2 nt s ->
   g_uni_rules g1 g2 (Transf2_uni_nt nt) (map g_uni_sf_lift2 s).
\end{coq}

Finally, \texttt {g\_uni} describes how to create a union grammar from two arbitrary source grammars. It uses the previous definitions to give values to each of the components of a new grammar definition.

\begin{coq}
Definition g_uni 
(non_terminal_1 non_terminal_2 terminal : Type)
(g1: cfg non_terminal_1 terminal) 
(g2: cfg non_terminal_2 terminal)
: (cfg g_uni_nt terminal):= 
   {| start_symbol:= Start_uni;
      rules:= g_uni_rules g1 g2;
      rules_finite:= g_uni_finite g1 g2 |}.
\end{coq}

Similar definitions were created to represent the concatenation of any two grammars and the closure of a grammar.

\subsection {Concatenation}

Given two arbitrary context-free grammars $g_1$ and $g_2$, the following definitions are used to construct $g_3$ such that $L(g_3)=L(g_1) \cdot L(g_2)$ (that is, the language generated by $g_3$ is the concatenation of the languages generated by $g_1$ and $g_2$).

\begin{coq}
Inductive g_cat_nt (non_terminal_1 non_terminal_2 terminal : Type): Type:=
| Start_cat
| Transf1_cat_nt: non_terminal_1 -> g_cat_nt
| Transf2_cat_nt: non_terminal_2 -> g_cat_nt.

Notation sf1:= (list (non_terminal_1 + terminal)).
Notation sf2:= (list (non_terminal_2 + terminal)).
Notation sfc:= (list (g_cat_nt + terminal)).

Definition g_cat_sf_lift1 (c: non_terminal_1 + terminal): 
g_cat_nt + terminal:=
  match c with
  | inl nt => inl (Transf1_cat_nt nt)
  | inr t  => inr t
  end.

Definition g_cat_sf_lift2 (c: non_terminal_2 + terminal): 
g_cat_nt + terminal:=
  match c with
  | inl nt => inl (Transf2_cat_nt nt)
  | inr t  => inr t
  end.

Inductive g_cat_rules 
(non_terminal_1 non_terminal_2 terminal : Type)
(g1: cfg non_terminal_1 terminal) 
(g2: cfg non_terminal_2 terminal)
: g_cat_nt -> sfc -> Prop :=
| New_cat: 
   g_cat_rules g1 g2 Start_cat 
   ([inl (Transf1_cat_nt (start_symbol g1))]++ 
    [inl (Transf2_cat_nt (start_symbol g2))])
| Lift1_cat: 
   forall nt s,
   rules g1 nt s ->
   g_cat_rules g1 g2 (Transf1_cat_nt nt) (map g_cat_sf_lift1 s)
| Lift2_cat: 
   forall nt s,
   rules g2 nt s -> 
   g_cat_rules g1 g2 (Transf2_cat_nt nt) (map g_cat_sf_lift2 s).

Definition g_cat 
(non_terminal_1 non_terminal_2 terminal : Type)
(g1: cfg non_terminal_1 terminal) 
(g2: cfg non_terminal_2 terminal)
: (cfg g_cat_nt terminal):= 
   {| start_symbol:= Start_cat;
      rules:= g_cat_rules g1 g2;
      rules_finite:= g_cat_finite g1 g2 |}.
\end{coq}

In this case, the new grammar (\texttt {g\_cat g1 g2}) is built in such a way that it has all the rules of \texttt {g1} and \texttt {g2}, plus a new rule that maps the new start symbol ({\texttt {Start\_cat}}) to the concatenation of the start symbols of the argument grammars (respectively \texttt {start\_symbol g1} and \texttt {start\_symbol g2}).

\subsection {Kleene Star}

Given an arbitrary context-free grammar $g_1$, the following definitions are used to construct $g_2$ such that $L(g_2)=(L(g_1))^*$ (that is, the language generated by $g_2$ is the reflexive and transitive concatenation (Kleene star) of the language generated by $g_1$).

\begin{coq}
Notation sfc:= (list (g_clo_nt + terminal)).

Inductive g_clo_nt (non_terminal : Type): Type :=
| Start_clo : g_clo_nt
| Transf_clo_nt : non_terminal -> g_clo_nt.

Definition g_clo_sf_lift (c: non_terminal + terminal): 
g_clo_nt + terminal:=
  match c with
  | inl nt => inl (Transf_clo_nt nt)
  | inr t  => inr t
  end.

Inductive g_clo_rules 
(non_terminal terminal : Type)
(g: cfg non_terminal terminal)
: g_clo_nt -> sfc -> Prop :=
| New1_clo: 
   g_clo_rules g Start_clo ([inl Start_clo] ++ 
   [inl (Transf_clo_nt (start_symbol g))])
| New2_clo: 
   g_clo_rules g Start_clo []
| Lift_clo: 
   forall nt: non_terminal,
   forall s: sf,
   rules g nt s ->
   g_clo_rules g (Transf_clo_nt nt) (map g_clo_sf_lift s).

Definition g_clo (g: cfg non_terminal terminal): 
(non_terminal terminal : Type)
(g: cfg g_clo_nt terminal):= 
 {| start_symbol:= Start_clo;
    rules:= g_clo_rules g;
    rules_finite:= g_clo_finite g |}.
\end{coq}

In this case, the new grammar (\texttt {g\_clo g}) is built in such a way that it has all the rules of \texttt {g}, plus two new rules: one that generates the empty string directly from the start symbol of \texttt {g\_clo g}, and another one that allows for the arbitrary concatenation of strings generated by \texttt {g}.

\subsection {Correctness and Completeness}

Although simple in their structure, it must be proved that the definitions \texttt {g\_uni}, \texttt {g\_cat} and \texttt {g\_clo} always produce the correct result. In other words, these definitions must be ``certified'', which is one of the main goals of formalization. In order to accomplish this, we must first state the theorems that capture the expected semantics of these definitions. Finally, we have to derive proofs of the correctness of these theorems. 

This can be done with a pair of theorems for each grammar definition: the first relates the output to the inputs, and the other one does the converse, providing assumptions about the inputs once an output is generated. This is necessary in order to guarantee that the definitions do only what one would expect, and no more.

In what follows, an informal statement is presented right before the corresponding Coq theorem. This is intended to abstract over the necessary mappings that occur in the Coq terms, due to the different types of non-terminals, sentential forms etc involved.

For concatenation, the following Coq statement expresses the result we want to prove (considering that $g_3$ is the concatenation of $g_1$ and $g_2$ and $S_3, S_1$ and $S_2$ are, respectively, the start symbols of $g_3, g_1$ and $g_2$): $$\forall g_1\,g_2,\,s_1,\,s_2, (S_1 \Rightarrow^*_{g_1} s_1) \land (S_2 \Rightarrow^*_{g_2} s_2) \rightarrow (S_3 \Rightarrow^*_{g_3} s_1s_2)$$

\begin{coq}
Theorem g_cat_correct:
forall g1: cfg non_terminal_1 terminal,
forall g2: cfg non_terminal_2 terminal,
forall s1: sf1,
forall s2: sf2,
generates g1 s1 /\ generates g2 s2 -> 
generates (g_cat g1 g2) ((map g_cat_sf_lift1 s1)++(map g_cat_sf_lift2 s2)).
\end{coq}

The above theorem states that if context-free grammars \texttt{g1} and \texttt{g2} generate, respectively, strings \texttt{s1} and \texttt{s2}, then the concatenation of these two grammars, according to the proposed algorithm, generates the concatenation of string \texttt{s1} with string \texttt{s2}. As mentioned before, the above theorem alone does not guarantee that \texttt{g\_cat} will not produce outputs other than the concatenation of its input strings. This idea is captured by the following complementary theorem: $$\forall s_3, (S_3 \Rightarrow^*_{g_3} s_3) \rightarrow \exists s_1,s_2\,|\,(s_3 = s_1 \cdot s_2) \land (S_1 \Rightarrow^*_{g_1} s_1) \land (S_2 \Rightarrow^*_{g_2} s_2)$$

For the converse of concatenation, the following Coq statement expresses the result we want to prove: 

\begin{coq}
Theorem g_cat_correct_inv:
forall g1: cfg non_terminal_1 terminal,
forall g2: cfg non_terminal_2 terminal,
forall s: sfc,
generates (g_cat g1 g2) s ->
s = [inl (start_symbol (g_cat g1 g2))] \/
exists s1: sf1,
exists s2: sf2,
s =(map g_cat_sf_lift1 s1)++(map g_cat_sf_lift2 s2) /\ 
generates g1 s1 /\ generates g2 s2.
\end{coq}

The idea here is to express that, if a string is generated by \texttt{g\_cat}, then it must only result from the concatenation of strings generated by the grammars combined by the definition. Together, these two theorems represent the semantics of the context-free grammar concatenation operation. The same ideas have been applied to the statement and proof of the following theorems, relative to the union and closure operations.

For union, we need to prove (considering that $g_3$ is the union of $g_1$ and $g_2$ and $S_3, S_1$ and $S_2$ are, respectively, the start symbols of $g_3, g_1$ and $g_2$): $$\forall g_1,\,g_2,\,s_1,\,s_2, (S_1 \Rightarrow^*_{g_1} s_1 \rightarrow S_3 \Rightarrow^*_{g_3} s_1) \land (S_2 \Rightarrow^*_{g_2} s_2 \rightarrow S_3 \Rightarrow^*_{g_3} s_2)$$

\noindent which translates in Coq into:

\begin{coq}
Theorem g_uni_correct:
forall g1: cfg non_terminal_1 terminal,
forall g2: cfg non_terminal_2 terminal,
forall s1: sf1,
forall s2: sf2,
(generates g1 s1 -> generates (g_uni g1 g2) (map g_uni_sf_lift1 s1)) 
/\
(generates g2 s2 -> generates (g_uni g1 g2) (map g_uni_sf_lift2 s2)).
\end{coq}

For the converse of union we have: $$\forall s_3, (S_3 \Rightarrow^*_{g_3} s_3) \rightarrow (S_1 \Rightarrow^*_{g_1} s_3) \lor (S_2 \Rightarrow^*_{g_2} s_3)$$

\begin{coq}
Theorem g_uni_correct_inv:
forall g1: cfg non_terminal_1 terminal,
forall g2: cfg non_terminal_2 terminal,
forall s: sfu,
generates (g_uni g1 g2) s ->
(s=[inl (start_symbol (g_uni g1 g2))]) \/
(exists s1: sf1, (s=(map g_uni_sf_lift1 s1) /\ generates g1 s1)) \/
(exists s2: sf2, (s=(map g_uni_sf_lift2 s2) /\ generates g2 s2)).
\end{coq}

For closure, we have (considering that $g_2$ is the Kleene star of $g_1$ and $S_2$ and $S_1$ are, respectively, the start symbols of $g_2$ and $g_1$): $$\forall g_1,\,s_1,\,s_2, (S_2 \Rightarrow^*_{g_2} \epsilon) \land ((S_2 \Rightarrow^*_{g_2} s_2) \land (S_1 \Rightarrow^*_{g_1} s_1) \rightarrow S_2 \Rightarrow^*_{g_2} s_2 \cdot s_1)$$

\noindent which translates in Coq into:
\begin{coq}
Theorem g_clo_correct:
forall g: cfg non_terminal terminal,
forall s: sf,
forall s': sfc,
generates (g_clo g) nil /\ (generates (g_clo g) s' /\ generates g s -> 
generates (g_clo g) (s'++ map g_clo_sf_lift s)).
\end{coq}

Finally: $$\forall s_2, (S_2 \Rightarrow^*_{g_2} s_2) \rightarrow (s_2 = \epsilon) \lor (\exists s_1,\,s_2'\,|\,(s_2 = s_2' \cdot s_1) \land (S_2 \Rightarrow^*_{g_2} s_2') \land (S_1 \Rightarrow^*_{g_1} s_1))$$

\begin{coq}
Theorem g_clo_correct_inv:
forall g: cfg non_terminal terminal,
forall s: sfc,
generates (g_clo g) s -> 
(s=[]) \/
(s=[inl (start_symbol (g_clo g))]) \/
(exists s': sfc, 
 exists s'': sf,
 generates (g_clo g) s' /\ generates g s'' /\  s=s' ++ map g_clo_sf_lift s'').
\end{coq}

The proofs of all the six main theorems have been completed (\texttt {g\_uni\_correct} and \texttt {g\_uni\_correct\_inv} for union, \texttt {g\_cat\_correct} and \texttt {g\_cat\_correct\_inv} for concatenation and \texttt {g\_clo\_correct} and \texttt {g\_clo\_correct\_inv} for closure). Most of them were obtained through induction over the predicate \texttt {derives} or one of its variants.

\subsection {Closure over Languages}

The previous results were all formulated over grammars, and it is desirable to obtain equivalent versions using languages instead. Thus, we have defined the union, concatenation and closure of arbitrary languages as follows:

\begin{coq}
Inductive l_uni (terminal : Type) (l1 l2: lang terminal): lang terminal:=
| l_uni_l1: forall s: sentence, l1 s -> l_uni l1 l2 s
| l_uni_l2: forall s: sentence, l2 s -> l_uni l1 l2 s.

Inductive l_cat (terminal : Type) (l1 l2: lang terminal): lang terminal:=
| l_cat_app: forall s1 s2: sentence, l1 s1 -> l2 s2 -> l_cat l1 l2 (s1 ++ s2).

Inductive l_clo (terminal : Type) (l: lang terminal): lang terminal:=
| l_clo_nil: l_clo l []
| l_clo_app: forall s1 s2: sentence, (l_clo l) s1 -> l s2 -> l_clo l (s1 ++ s2). 
\end{coq}

With these definitions, it is immediate to prove that the operations of union, concatenation and closure are correct, including the converse versions. However, it remains to be proved that the newly generated languages are also context-free, which leads to the following theorems:

\begin{coq}
Theorem l_uni_is_cfl:
forall l1 l2: lang terminal,
cfl l1 -> cfl l2 -> cfl (l_uni l1 l2).

Theorem l_cat_is_cfl:
forall l1 l2: lang terminal,
cfl l1 -> cfl l2 -> cfl (l_cat l1 l2).

Theorem l_clo_is_cfl:
forall l: lang terminal,
cfl l -> cfl (l_clo l).
\end{coq}

In all cases, the proofs obtained rely on (i) the existence of context-free grammars that generated the original languages, a direct consequence of the definition of \texttt {cfl} and (ii) the results that were previously proved for context-free grammars.

\section {Simplification}
\label {sec-simplification}

The definition of a context-free grammar, and also the operations defined in the previous section, allow for the inclusion of symbols and rules that might not contribute to the language being generated. Besides that, context-free grammars might also contain rules that can be substituted by equivalent smaller and simpler ones. Unit rules, for example, do not expand sentential forms (instead, they just rename the symbols in them) and empty rules can cause them to contract. Although the appropriate use of these features can be important for human communication in some situations, this is not the general case, since it leads to grammars that have more symbols and rules than necessary, making difficult its comprehension and manipulation. Thus, simplification is an important operation on context-free grammars.

Let $G$ be a context-free grammar, $L(G)$ the language generated by this grammar and $\epsilon$ the empty string. Different authors use different terminology when presenting simplification results for context-free grammars. In what follows, we adopt the terminology and definitions of \cite {sudkamp-2006}. 

Context-free grammar simplification comprises the manipulation of rules and symbols, as described below:

\begin {enumerate}
\item \label {empty}
An \emph {empty rule} $r \in P$ is a rule whose right-hand side $\beta$ is empty (e.g. $X \rightarrow \epsilon$). We formalize that for all $G$, there exists $G'$ such that $L(G)=L(G')$ and $G'$ has no empty rules, except for a single rule $S \rightarrow \epsilon$ if $\epsilon \in L(G)$; in this case, $S$ (the initial symbol of $G'$) does not appear on the right-hand side of any rule in $G'$;

\item \label {unit} 
A \emph {unit rule} $r \in P$ is a rule whose right-hand side $\beta$ contains a single non-terminal symbol (e.g. $X \rightarrow Y$). We formalize that for all $G$, there exists $G'$ such that $L(G)=L(G')$ and $G'$ has no unit rules;

\item \label {useless} 
$s \in V$ is \emph {useful} (\cite {sudkamp-2006}, p. 116) if it is possible to derive a sentence from it using the rules of the grammar. Otherwise, $s$ is called an \emph {useless symbol}. A useful symbol $s$ is one such that $s \Rightarrow^* \omega$, with $\omega \in \Sigma^*$. Naturally, this definition concerns mainly non-terminals, as terminals are trivially useful. We formalize that, for all $G$ such that $L(G) \neq \emptyset$, there exists $G'$ such that $L(G)=L(G')$ and $G'$ has no useless symbols;

\item \label {inaccessible} 
$s \in V$ is \emph {accessible} (\cite {sudkamp-2006}, p. 119) if it is part of at least one string generated from the root symbol of the grammar. Otherwise, it is called an \emph {inaccessible symbol}. An accessible symbol $s$ is one such that $S \Rightarrow^* \alpha s\beta$, with $\alpha, \beta \in V^*$. We formalize that for all $G$, there exists $G'$ such that $L(G)=L(G')$ and $G'$ has no inaccessible symbols.
\end {enumerate}

Finally, we formalize a unification result: that for all $G$, if $G$ is non-empty, then there exists $G'$ such that $L(G)=L(G')$ and $G'$ has no empty rules (except for one, if $G$ generates the empty string), no unit rules, no useless symbols, no inaccessible symbols and the start symbol of $G'$ does not appear on the right-hand side of any other rule of $G'$.

In all these four cases and the five grammars that are discussed next (namely \texttt {g\_emp}, \texttt {g\_emp'}, \texttt {g\_unit}, \texttt {g\_use} and \texttt {g\_acc}), the proof of \texttt {rules\_finite} is based on the proof of the correspondent predicate for the argument grammar. Thus, all new grammars satisfy the \texttt {cfg} specification and are finite as well.

\subsection {Empty rules}
Result (\ref {empty}) is achieved in two steps. First, the idea of a \emph {nullable} (\cite {sudkamp-2006}, p. 107) symbol was represented by the definition \texttt {empty}: 

\begin{coq}
Definition empty 
(g: cfg terminal _) (s: non_terminal + terminal): Prop:=
derives g [s] [].
\end{coq}

Notation \texttt {sf'} represents a sentential form that is constructed with elements of \texttt {non\_terminal'} and \texttt {terminal}. Definition \texttt {symbol\_lift} maps a pair of type \texttt {(non\_terminal + terminal)} into a pair of type \texttt {(non\_terminal' + terminal)} by replacing each \texttt {non\_terminal} with the corresponding \texttt {non\_terminal'}:

\begin{coq}
Inductive non_terminal': Type:=
| Lift_nt: non_terminal -> non_terminal'
| New_ss. 

Notation sf' := (list (non_terminal' + terminal)).

Definition symbol_lift 
(s: non_terminal + terminal): non_terminal' + terminal:=
match s with
| inr t => inr t
| inl n => inl (Lift_nt n)
end.
\end{coq}

With these, a new grammar \texttt {g\_emp g} has been created, such that the language generated by it matches the language generated by the original grammar (\texttt {g}), except for the empty string. Predicate \texttt {g\_emp\_rules} states that every non-empty rule of \texttt {g} is also a rule of \texttt {g\_emp g}, and also adds new rules to \texttt {g\_emp g} where every possible combination of nullable non-terminal symbols that appears on the right-hand side of a rule of \texttt {g} is removed, as long as the resulting right-hand side is not empty. Finally, it adds a rule that maps a new symbol, the start symbol of the new grammar (\texttt {New\_ss}), to the start symbol of the original grammar. For this reason, the new type \texttt {non\_terminal'} has been defined. The motivation for introducing a new start symbol at this point is to be able to prove that the start symbol does not appear in the right-hand side of any rule of the new grammar, a result that will be important in future developments.

\begin{coq}
Inductive g_emp_rules 
(non_terminal terminal : Type) 
(g: cfg non_terminal terminal)
: non_terminal' -> sf' -> Prop :=
| Lift_direct : 
    forall left: non_terminal,
    forall right: sf,
    right <> [] -> rules g left right ->
    g_emp_rules g (Lift_nt left) (map symbol_lift right)
| Lift_indirect:
    forall left: non_terminal,
    forall right: sf,
    g_emp_rules g (Lift_nt left) (map symbol_lift right)->
    forall s1 s2: sf, 
    forall s: non_terminal,
    right = s1 ++ (inl s) :: s2 ->
    empty g (inl s) ->
    s1 ++ s2 <> [] ->
    g_emp_rules g (Lift_nt left) (map symbol_lift (s1 ++ s2))
| Lift_start_emp: 
    g_emp_rules g New_ss [inl (Lift_nt (start_symbol g))]. 
	   
Definition g_emp 
(non_terminal terminal : Type)
(g: cfg non_terminal terminal)
: cfg non_terminal' terminal := 
  {| start_symbol:= New_ss;
     rules:= g_emp_rules g;
     rules_finite:= g_emp_finite g |}.
\end{coq}

Suppose, for example, that $X, A, B, C$ are non-terminals, of which $A, B$ and $C$ are nullable, $a, b$ and $c$ are terminals and $X \rightarrow aAbBcC$ is a rule of \texttt {g}. Then, the above definitions assert that $X \rightarrow aAbBcC$ is a rule of \texttt {g\_emp g}, and also:

\begin {itemize}
\item $X \rightarrow aAbBc$;
\item $X \rightarrow abBcC$;
\item $X \rightarrow aAbcC$;
\item $X \rightarrow aAbc$;
\item $X \rightarrow abBc$;
\item $X \rightarrow abcC$;
\item $X \rightarrow abc$.
\end {itemize}

Observe that grammar \texttt {g\_emp g} does not generate the empty string. The second step, thus, was to define \texttt {g\_emp' g}, such that \texttt {g\_emp' g} generates the empty string if \texttt {g} generates the empty string. This was done by stating that every rule from \texttt {g\_emp g} is also a rule of \texttt {g\_emp' g} and also by adding a new rule that allow \texttt {g\_emp' g} to generate the empty string directly if necessary. 

\begin{coq}
Inductive g_emp'_rules 
(non_terminal terminal : Type)
(g: cfg non_terminal terminal)
: non_terminal' non_terminal -> sf' -> Prop :=
| Lift_all:
   forall left: non_terminal' _,
   forall right: sf',
   rules (g_emp g) left right -> g_emp'_rules g left right
| Lift_empty:
   empty g (inl (start_symbol g)) -> g_emp'_rules g (start_symbol (g_emp g)) [].

Definition g_emp' 
(non_terminal terminal : Type)
(g: cfg non_terminal terminal)
: cfg (non_terminal' _) terminal := 
  {| start_symbol:= New_ss _;
     rules:= g_emp'_rules g;
     rules_finite:= g_emp'_finite g |}.
\end{coq}

Note that the generation of the empty string by \texttt {g\_emp' g} depends on \texttt {g} generating the empty string. 

The proof of the correctness of these definitions is achieved through the following theorem:

\begin{coq}
Theorem g_emp'_correct: 
forall g: cfg non_terminal terminal,
g_equiv (g_emp' g) g /\
(generates_empty g -> has_one_empty_rule (g_emp' g)) /\ 
(~ generates_empty g -> has_no_empty_rules (g_emp' g)) /\
start_symbol_not_in_rhs (g_emp' g).
\end{coq}

Four auxiliary predicates have been used in this statement: \texttt {g\_equiv} (introduced in Section \ref {sub-languages}) for two context-free grammars that generate the same language, \texttt {generates\_empty} for a grammar whose language includes the empty string, \texttt {has\_one\_empty\_rule} for a grammar that has an empty rule whose left-hand side is the initial symbol, and all other rules are not empty and \texttt {has\_no\_empty\_rules} for a grammar that has no empty rules at all.

The definition of \texttt {g\_equiv}, when applied to this theorem, yields:

\begin{coq}
forall s: sentence,
produces (g_emp' g) s <-> produces g s.
\end{coq}

For the $\rightarrow$ part, the strategy is to prove that for every rule $left \rightarrow_{g\_emp'} right$, either $left \rightarrow_{g} right$ is a rule of \texttt {g} or $left \Rightarrow^*_{g} right$. For the $\leftarrow$ part, the strategy is a more complicated one, and involves induction over the number of derivation steps in \texttt {g}.

\subsection {Unit rules}

For result (\ref {unit}), definition \texttt {unit} expresses the relation between any two non-terminal symbols $X$ and $Y$, and is true when $X \Rightarrow^* Y$ (\cite {sudkamp-2006}, p. 114).

\begin{coq}
Inductive unit 
(terminal non_terminal : Type)
(g: cfg terminal non_terminal) 
(a: non_terminal)
: non_terminal -> Prop:=
| unit_rule: 
   forall (b: non_terminal),
   rules g a [inl b] -> unit g a b
| unit_trans: 
   forall b c: non_terminal,
   unit g a b -> unit g b c -> unit g a c.
\end{coq}

Grammar \texttt {g\_unit g} represents the grammar that is equivalent to \texttt {g}, except that the unit rules of the latter have been substituted by others, non-unit rules, that produce the same results in terms of the generated language. The idea is that \texttt {g\_unit g} has all non-unit rules of \texttt {g}, plus new rules that are created by anticipating the possible application of unit rules in \texttt {g}, as informed by \texttt {g\_unit}.

\begin{coq}
Inductive g_unit_rules 
(terminal non_terminal : Type)
(g: cfg non_terminal terminal)
: non_terminal -> sf -> Prop :=
| Lift_direct' : 
   forall left: non_terminal,
   forall right: sf,
   (forall r: non_terminal, right <> [inl r]) -> 
   rules g left right ->
   g_unit_rules g left right
| Lift_indirect':
   forall a b: non_terminal,
   unit g a b ->
   forall right: sf,
   rules g b right ->  
   (forall c: non_terminal, right <> [inl c]) -> 
   g_unit_rules g a right.

Definition g_unit 
(terminal non_terminal : Type)
(g: cfg non_terminal terminal)
: cfg non_terminal terminal := 
  {| start_symbol:= start_symbol g;
     rules:= g_unit_rules g;
     rules_finite:= g_unit_finite g |}.
\end{coq}

Finally, the correctness of \texttt {g\_unit} comes from the following theorem:

\begin{coq}
Theorem g_unit_correct: 
forall g: cfg non_terminal terminal,
g_equiv (g_unit g) g /\ has_no_unit_rules (g_unit g).
\end{coq}

The predicate \texttt {has\_no\_unit\_rules} states that the argument grammar has no unit rules at all.

Similar to the previous case, for the $\rightarrow$ part of the \texttt {g\_equiv (g\_unit g) g} proof, the strategy adopted is to prove that for every rule $left \rightarrow_{g\_unit} right$ of (\texttt {g\_unit g}), either $left \rightarrow_{g} right$ is a rule of \texttt {g} or $left \Rightarrow^*_{g} right$. For the $\leftarrow$ part, the strategy is also a more complicated one, and involves induction over a predicate that is equivalent to \emph {derives} (\emph {derives3}), but generates the sentence directly without considering the application of a sequence of rules, which allows one to abstract the application of unit rules in \texttt {g}.

\subsection {Useless symbols}
For result (\ref {useless}), the idea of a useful symbol is captured by the definition \texttt {useful}:

\begin{coq}
Definition useful 
(terminal non_terminal : Type)
(g: cfg non_terminal terminal) 
(s: non_terminal + terminal): Prop:=
match s with
| inr t => True
| inl n => exists s: sentence, derives g [inl n] (map term_lift s)
end.
\end{coq}

The removal of useless symbols comprises, first, the identification of useless symbols in the grammar and, second, the elimination of the rules that use them. Definition \texttt {g\_use\_rules} selects, from the original grammar, only the rules that do not contain useless symbols. The new grammar, without useless symbols, can then be defined as in \texttt {g\_use}:

\begin{coq}
Inductive g_use_rules 
(terminal non_terminal : Type)
(g: cfg non_terminal terminal) 
: non_terminal -> sf -> Prop :=
| Lift_use : 
   forall left: non_terminal,
   forall right: sf,
   rules g left right ->
   useful g (inl left) ->
   (forall s: non_terminal + terminal, In s right -> useful g s) -> 
   g_use_rules g left right.

Definition g_use 
(terminal non_terminal : Type)
(g: cfg non_terminal terminal) 
: cfg non_terminal terminal:= 
  {| start_symbol:= start_symbol g;
     rules:= g_use_rules g;
     rules_finite:= g_use_finite g |}.			 
\end{coq}

The \texttt {g\_use} definition, of course, can only be used if the language generated by the original grammar is not empty, that is, if the start symbol of the original grammar is useful. If it were useless then it would be impossible to assign a root to the grammar and the language would be empty. The correctness of the useless symbol elimination operation can be certified by proving theorem \texttt {g\_use\_correct}, which states that every context-free grammar whose start symbol is useful generates a language that can also be generated by an equivalent context-free grammar whose symbols are all useful.

\begin{coq}
Theorem g_use_correct: 
forall g: cfg non_terminal terminal,
non_empty g -> g_equiv (g_use g) g /\ has_no_useless_symbols (g_use g). 
\end{coq}

The predicates \texttt {non\_empty}, and \texttt {has\_no\_useless\_symbols} used above assert, respectively, that grammar \texttt {g} generates a language that contains at least one string (which in turn may or may not be empty) and the grammar has no useless symbols at all.

The $\rightarrow$ part of the \texttt {g\_equiv} proof is straightforward, since every rule of \texttt {g\_use} is also a rule of \texttt {g}. For the converse, it is necessary to show that every symbol used a the derivation of \texttt {g} is useful, and thus the rules used in this derivation also appear in \texttt {g\_use}.

\subsection {Inaccessible symbols}

Result (\ref {inaccessible}) is similar to the previous case, and definition \texttt {accessible} has been used to represent accessible symbols in context-free grammars.

\begin{coq}
Definition accessible 
(terminal non_terminal : Type) 
(g : cfg non_terminal terminal)
(s: non_terminal + terminal): Prop:=
exists s1 s2: sf, derives g [inl (start_symbol g)] (s1++s::s2).
\end{coq}

Definition \texttt {g\_acc\_rules} selects, from the original grammar, only the rules that do not contain inaccessible symbols. Definition \texttt {g\_acc} represents a grammar whose inaccessible symbols have been removed:

\begin{coq}
Inductive g_acc_rules 
(terminal non_terminal : Type) 
(g : cfg non_terminal terminal)
: non_terminal -> sf -> Prop :=
| Lift_acc : forall left: non_terminal,
   forall right: sf,
   rules g left right -> accessible g (inl left) -> g_acc_rules g left right.

Definition g_acc 
(terminal non_terminal : Type) 
(g : cfg non_terminal terminal)
: cfg non_terminal terminal := 
  {| start_symbol:= start_symbol g;
     rules:= g_acc_rules g;
     rules_finite:= g_acc_finite g |}.
\end{coq}

The correctness of the inaccessible symbol elimination operation can be certified by proving theorem \texttt {g\_acc\_correct}, which states that every context-free grammar generates a language that can also be generated by an equivalent context-free grammar where symbols are all accessible.

\begin{coq}
Theorem g_acc_correct: 
forall g: cfg non_terminal terminal,
g_equiv (g_acc g) g /\ has_no_inaccessible_symbols (g_acc g).  
\end{coq}

In a way similar to \texttt {has\_no\_useless\_symbols}, the absence of inaccessible symbols in a grammar is expressed by predicate \texttt {has\_no\_inaccessible\_symbols} used above.

Similar to the previous case, the $\rightarrow$ part of the \texttt {g\_equiv} proof is also straightforward, since every rule of \texttt {g\_acc} is also a rule of \texttt {g}. For the converse, it is necessary to show that every symbol used in the derivation of \texttt {g} is accessible, and thus the rules used in this derivation also appear in \texttt {g\_acc}.

\subsection {Unification}

If one wants to obtain a new grammar simultaneously free of empty and unit rules, and of useless and inaccessible symbols, it is not enough to consider the previous independent results: it is necessary to establish a suitable order to apply these simplifications, in order to guarantee that the final result satisfies all desired conditions. Then, it is necessary to prove that the claims do hold.

For the order, we should start with (i) the elimination of empty rules, followed by (ii) the elimination of unit rules. The reason for this is that (i) might introduce new unit rules in the grammar, and (ii) will surely not introduce empty rules, as long as original grammar is free of them (except for $S \rightarrow \epsilon$, in which case $S$, the initial symbol of the grammar, must not appear on the right-hand side of any rule). Then, elimination of useless and inaccessible symbols (in either order) is the right thing to do, since they only remove rules from the original grammar (which is specially important because they do not introduce new empty or unit rules).

The formalization of this result is captured in the following theorem, which represents the main result of this section:

\begin{coq}
Theorem g_simpl:
forall g: cfg non_terminal terminal,
non_empty g ->
exists g': cfg (non_terminal' non_terminal) terminal,
g_equiv g' g /\
has_no_inaccessible_symbols g' /\
has_no_useless_symbols g' /\
(generates_empty g -> has_one_empty_rule g') /\ 
(~ generates_empty g -> has_no_empty_rules g') /\
 has_no_unit_rules g' /\
 start_symbol_not_in_rhs g'.
\end{coq}

Hypothesis \texttt {non\_empty g} is necessary in order to allow for the elimination of useless symbols. The predicate \texttt {start\_symbol\_not\_in\_rhs} states that the start symbol does not appear in the right-hand side of any rule of the argument grammar.

The proof of \texttt {g\_simpl} demands auxiliary lemmas to prove that the characteristics of the initial transformations are preserved by the following ones. For example, that all of the unit rules elimination, useless symbol elimination and inaccessible symbol elimination operations preserve the characteristics of the empty rules elimination operation.

\section {Chomsky Normal Form}
\label {sec-cnf}

The Chomsky Normal Form (CNF) theorem asserts: 
$$\forall G=(V,\Sigma,P,S),\;\exists G'=(V',\Sigma,P',S')\;|$$
$$L(G)=L(G') \land \forall (\alpha \rightarrow_{G'} \beta) \in P', (\beta \in \Sigma) \lor (\beta \in N\cdot N)$$

That is, every context-free grammar can be converted to an equivalent one whose rules have only one terminal symbol or two non-terminal symbols in the right-hand side. Naturally, this is valid only if $G$ does not generate the empty string. If this is the case, then the grammar that has this format, plus a single rule $S' \rightarrow_{G} \epsilon$, is also considered to be in the Chomsky Normal Form, and generates the original language, including the empty string. It can also be assured that in either case the start symbol of $G'$ does not appear on the right-hand side of any rule of $G'$.

The existence of a CNF can be used for a variety of purposes, including to prove that there is an algorithm to decide whether an arbitrary context-free language accepts an arbitrary string, and to test if a language is not context-free (using the Pumping Lemma for context-free languages, which can be proved with the help of CNF grammars).

The idea of mapping $G$ into $G'$ consists of creating a finite number of new non-terminal symbols and new rules, in the following ways:

\begin {enumerate}
\item For every terminal symbol $\sigma$ that appears in the right-hand side of a rule $r = \alpha \rightarrow_{G} \beta_1 \cdot \sigma \cdot \beta_2$ of $G$, create a new non-terminal symbol $[\sigma]$, a new rule $[\sigma] \rightarrow_{G'} \sigma$ and substitute $\sigma$ for $[\sigma]$ in $r$; 
\item For every rule $r = \alpha \rightarrow_G N_1 N_2 \cdots N_k$ of $G$, where $N_i$ are all non-terminals, create a new set of non-terminals and a new set of rules such that: 
  \begin {eqnarray*}
  \alpha & \rightarrow_{G'} & N_1 \lbrack N_2 \cdots N_k \rbrack, \\
  \lbrack N_2 \cdots N_k \rbrack & \rightarrow_{G'} & N_2 \lbrack N_3 \cdots N_k \rbrack, \\
  & \cdots & \\
  \lbrack N_{k-2}N_{k-1}N_k \rbrack & \rightarrow_{G'} & N_{k-2} \lbrack N_{k-1}N_k \rbrack, \\
  \lbrack N_{k-1}N_k \rbrack & \rightarrow_{G'} & N_{k-1} N_k
  \end {eqnarray*}
\end {enumerate}

Case (i) substitutes all terminal for non-terminal symbols. Case (ii) splits rules that have three or more non-terminal symbols on the right-hand side by a set of rules that has only two non-terminal symbols in the right-and side. Both changes preserve the language of the original grammar. 

As an example, consider $G=(\{S',X,Y,Z,a,b,c\},\{a,b,c\},P,S')$ with $P$ equal to:

\begin{eqnarray*}
\{ S' & \rightarrow & XYZd, \\
   X  & \rightarrow & a, \\
   Y  & \rightarrow & b, \\
   Z  & \rightarrow & c, \}
\end{eqnarray*}

The CNF grammar $G'$, equivalent to $G$, would then be the one with the following set of rules:

\begin{eqnarray*}
\{ S'    & \rightarrow & X\lbrack YZd\rbrack, \\
   \lbrack YZd\rbrack & \rightarrow & Y\lbrack Zd\rbrack, \\
   \lbrack Zd\rbrack  & \rightarrow & Z\lbrack d\rbrack, \\
   \lbrack d\rbrack   & \rightarrow & d, \\
      X  & \rightarrow & a, \\
      Y  & \rightarrow & b, \\
      Z  & \rightarrow & c, \}
\end{eqnarray*}

These ideas  are captured by the following definitions. The non-terminals of the new grammar \texttt {g\_cnf g} are represented by the type \texttt {non\_terminal'}. Its elements are associated with sentential forms of \texttt {g} via the constructor \texttt {Lift\_r}:

\begin{coq}
Inductive non_terminal' (non_terminal terminal : Type): Type:=
| Lift_r: sf -> non_terminal'.

Notation sf':= (list (non_terminal' + terminal)).
Notation term_lift:= ((terminal_lift non_terminal) terminal).
\end{coq}

The function \texttt {symbol\_lift}, presented below, maps sentential forms of \texttt {g} into sentential forms of \texttt {g\_cnf g}:

\begin{coq}
Definition symbol_lift (s: non_terminal + terminal)
: non_terminal' + terminal:=
match s with
| inr t => inr t
| inl n => inl (Lift_r [inl n])
end.
\end{coq}

The rules of \texttt {g\_cnf g} and \texttt {g\_cnf g} itself are defined as:

\begin{coq}
Inductive g_cnf_rules
(non_terminal terminal : Type)
(g: cfg non_terminal terminal)
: non_terminal' -> sf' -> Prop:=
| Lift_cnf_t:  
   forall t: terminal,
   forall left: non_terminal,
   forall s1 s2: sf,
   rules g left (s1++[inr t]++s2) ->
   g_cnf_rules g (Lift_r [inr t]) [inr t]
| Lift_cnf_1:  
   forall left: non_terminal,
   forall t: terminal,
   rules g left [inr t] ->
   g_cnf_rules g (Lift_r [inl left]) [inr t]
| Lift_cnf_2:  
   forall left: non_terminal,
   forall s1 s2: symbol,
   forall beta: sf,
   rules g left (s1 :: s2 :: beta) ->
   g_cnf_rules g (Lift_r [inl left]) 
   [inl (Lift_r [s1]); inl (Lift_r (s2 :: beta))]
| Lift_cnf_3:  
   forall left: sf,
   forall s1 s2 s3: symbol,
   forall beta: sf,
   g_cnf_rules g (Lift_r left) 
   [inl (Lift_r [s1]); inl (Lift_r (s2 :: s3 :: beta))] ->
   g_cnf_rules g (Lift_r (s2 :: s3 :: beta)) 
   [inl (Lift_r [s2]); inl (Lift_r (s3 :: beta))].

Definition g_cnf 
(non_terminal terminal : Type)
(g: cfg non_terminal terminal)
: cfg non_terminal' terminal := 
  {| start_symbol:= Lift_r [inl (start_symbol g)];
     rules:= g_cnf_rules g;
     rules_finite:= g_cnf_finite g |}.
\end{coq}

Next, we prove that \texttt {g\_cnf g} is equivalent to \texttt {g}. It should be noted, however, that the set of rules defined above do not generate the empty string. If this is the case, the definitions below define a new grammar \texttt {g\_cnf'} that adds a new rule that generates the empty string:

\begin{coq}
Inductive g_cnf'_rules 
(non_terminal terminal : Type)
(g: cfg non_terminal terminal)
: non_terminal' -> sf' -> Prop:=
| Lift_cnf'_all: 
   forall left: non_terminal',
   forall right: sf',
   g_cnf_rules g left right ->
   g_cnf'_rules g left right
| Lift_cnf'_new: 
   g_cnf'_rules g (start_symbol (g_cnf g)) [].
	   
Definition g_cnf' 
(non_terminal terminal : Type)
(g: cfg non_terminal terminal)
: cfg non_terminal' terminal:= 
  {| start_symbol:= start_symbol (g_cnf g);
     rules:= g_cnf'_rules g;
     rules_finite:= g_cnf'_finite g |}.
\end{coq}

The statement of the CNF theorem can then be presented as:

\begin{coq}
Theorem g_cnf_final:
forall g: cfg non_terminal terminal,
(produces_empty g \/ ~ produces_empty g) /\ 
(produces_non_empty g \/ ~ produces_non_empty g) ->
exists g': cfg non_terminal' terminal, 
g_equiv g' g /\ 
(is_cnf g' \/ is_cnf_with_empty_rule g').
\end{coq}

The predicates used above assert that the argument grammar:

\begin {itemize}
\item produces the empty string (\texttt {produces\_empty});
\item does not produce the empty string (\texttt {produces\_non\_empty});
\item is in the Chomsky Normal Form (\texttt {is\_cnf});
\item is in the Chomsky Normal Form and has a single empty rule with the start symbol in the left-hand side (\texttt {is\_cnf\_with\_empty\_rule}).
\end {itemize}

The proof of this theorem requires, among other things, that the original grammar is first simplified according to the results discussed in the previous section. For the $\leftarrow$ part of \texttt {g\_equiv}, the strategy adopted is to prove that for every rule $left \rightarrow right$ of (\texttt {g}), either $left \rightarrow right$ is a rule of \texttt {g\_cnf g} or $left \Rightarrow^* right$ in \texttt {g\_cnf g}. 

For the $\rightarrow$ part, that is, $(s_1 \Rightarrow^*_{g\_cnf g} s_2) \rightarrow (s_1 \Rightarrow^*_{g} s_2)$, it is enough to note that the sentential forms of \texttt {g} are embedded in the sentential forms of \texttt {g\_cnf g}, specifically in the arguments of the constructor \texttt {Lift\_r} of \texttt {non\_terminal'}. Thus, a simple extraction mechanism allows the implication to be proved by induction on the structure of the sentential form $s_1$.

Using the previous example, suppose we have: $X\lbrack YZd\rbrack \Rightarrow^*_{g\_cnf g} abcd$, which would be represented in our formalization as:

\begin{coq}
derives (g_cnf g) [inl X] ++ [inl (Lift_r ([inl Y; inl Z; inr d]))] 
(map (@symbol_lift _ _) (map term_lift [inr a; inr b; inr c; inr d]))
\end{coq}

The extraction mechanism, applied to this case, would yield:

\begin{coq}
derives g [inl X; inl Y; inl Z; inr d] (map term_lift [inr a; inr b; inr c; inr d])
\end{coq}

\noindent which is exactly the expected result $(XYZd \Rightarrow^*_g abcd)$.

\section {Related Work}
\label {sec-related}
Context-free language theory formalization is a relatively new area of research, with some results already obtained with a diversity of proof assistants, including Coq, HOL4 and Agda. Most of the effort started in 2010 and have been devoted to the certification and validation of parser generators. Examples of this are the works of Koprowski and Binsztok (using Coq, \cite {koprowski-2010}), Ridge (using HOL4, \cite {ridge-2011}), Jourdan, Pottier and Leroy (using Coq, \cite {jourdan-2012}) and, more recently, Firsov and Uustalu (in Agda, \cite {firsov-2014}). 

On the more theoretical side, on which the present work should be considered, Norrish and Barthwal published on general context-free language theory formalization using the using HOL4 proof assistant \cite {barthwal-norrish-2010a,barthwal-norrish-2010b,barthwal-norrish-2013}, including the existence of normal forms for grammars, pushdown automata and closure properties. Recently, Firsov and Uustalu proved the existence of a Chomsky Normal Form grammar for every general context-free grammar, using the Agda proof assistant \cite {firsov-2015}.

It can thus be noted that so far apparently no formalization has been done in Coq for results not related directly to parsing and parser verification (except in HOL4 and Agda), and that this constitutes an important motivation for the present work, mainly due to the increasing usage and importance of Coq in different areas and communities. Specifically, the formalization done by Norrish and Barthwal in HOL4 is quite comprehensive and extends our work with the Greibach Normal Form and pushdown automata and its relation to context-free grammars. It does not include, however, a proof of either the decidability of the membership problem or the Pumping Lemma for context-free languages, which are objectives of the present work. The formalization by Firsov and Uustalu in Agda comprises basically the existence of a Chomsky Normal Form, and formalizes the elimination of empty and unit rules, but not elimination of useless and inaccessible symbols.

\section {Conclusions}
\label {sec-conclusions}

All important objects related with context-free grammars have been properly represented and different grammar manipulation strategies were formalized. Proofs of their correctness were successfully constructed. The proofs of all lemmas and theorems presented in this article have been formalized in Coq and comprise approximately 18,000 lines of scripts. This number can be explained for the following reasons:

\begin {enumerate}
\item The style adopted for writing the scripts: for the sake of clarity, each tactic is placed in its own line, despite the possibility of combining several tactics in the same line. Also, bullets (for structuring the code) were used as much as possible and the sequence tactical (using the semicolon symbol) was avoided at all. This duplicates parts of the code but has the advantage of keeping the static structure of the script related to its dynamic behaviour, which favors legibility and maintenance.
\item The formalization includes not only the main theorems described here, but also an extensive library of other fundamental and auxiliary lemmas on context-free grammars and derivations, which have been used to obtain the main results presented here, were used in the previously obtained results and will be used in future developments.

The results presented in this paper are fundamental to context-free language theory. They create an adequate framework in which to pursue further results, including a proof of the decidability of the membership problem and a proof of the Pumping Lemma for context-free languages.
\end {enumerate}

\section*{Bibliography}
\bibliographystyle {elsarticle-harv}
\bibliography {article}

\begin{thebibliography}{13}
\expandafter\ifx\csname natexlab\endcsname\relax\def\natexlab#1{#1}\fi
\expandafter\ifx\csname url\endcsname\relax
  \def\url#1{\texttt{#1}}\fi
\expandafter\ifx\csname urlprefix\endcsname\relax\def\urlprefix{URL }\fi

\bibitem[{Barthwal and Norrish(2010{\natexlab{a}})}]{barthwal-norrish-2010a}
Barthwal, A., Norrish, M., August 2010{\natexlab{a}}. A formalisation of the
  normal forms of context-free grammars in {HOL4}. In: Dawar, A., Veith, H.
  (Eds.), Computer Science Logic, 24th International Workshop, CSL 2010, 19th
  Annual Conference of the EACSL, Brno, Czech Republic, August 23--27, 2010.
  Proceedings. Vol. 6247 of Lecture Notes in Computer Science. Springer, pp.
  95--109.

\bibitem[{Barthwal and Norrish(2010{\natexlab{b}})}]{barthwal-norrish-2010b}
Barthwal, A., Norrish, M., 2010{\natexlab{b}}. Mechanisation of {PDA} and
  grammar equivalence for context-free languages. In: Dawar, A., de~Queiroz, R.
  J. G.~B. (Eds.), Logic, Language, Information and Computation, 17th
  International Workshop, {WoLLIC}~2010. Vol. 6188 of Lecture Notes in Computer
  Science. pp. 125--135.

\bibitem[{Barthwal and Norrish(2014)}]{barthwal-norrish-2013}
Barthwal, A., Norrish, M., 2014. A mechanisation of some context-free language
  theory in {HOL4}. Journal of Computer and System Sciences (WoLLIC 2010
  Special Issue, A. Dawar and R. de Queiroz, eds.) 80~(2), 346 -- 362.
\newline\urlprefix\url{http://www.sciencedirect.com/science/article/pii/S0022000013000925}

\bibitem[{Bertot and Castéran(2004)}]{bertot-2004}
Bertot, Y., Castéran, P., 2004. Interactive Theorem Proving and Program
  Development. Springer.

\bibitem[{Firsov and Uustalu(2014)}]{firsov-2014}
Firsov, D., Uustalu, T., 2014. Certified \{CYK\} parsing of context-free
  languages. Journal of Logical and Algebraic Methods in Programming
  83~(5–6), 459 -- 468, the 24th Nordic Workshop on Programming Theory (NWPT
  2012).
\newline\urlprefix\url{http://www.sciencedirect.com/science/article/pii/S2352220814000601}

\bibitem[{Firsov and Uustalu(2015)}]{firsov-2015}
Firsov, D., Uustalu, T., 2015. Certified normalization of context-free
  grammars. In: Proceedings of the 2015 Conference on Certified Programs and
  Proofs. CPP '15. ACM, New York, NY, USA, pp. 167--174.
\newline\urlprefix\url{http://doi.acm.org/10.1145/2676724.2693177}

\bibitem[{Jourdan et~al.(2012)Jourdan, Pottier, and Leroy}]{jourdan-2012}
Jourdan, J.-H., Pottier, F., Leroy, X., 2012. Validating {L}{R}(1) parsers. In:
  Proceedings of the 21st European Conference on Programming Languages and
  Systems. ESOP'12. Springer-Verlag, Berlin, Heidelberg, pp. 397--416.
\newline\urlprefix\url{http://dx.doi.org/10.1007/978-3-642-28869-2_20}

\bibitem[{Koprowski and Binsztok(2010)}]{koprowski-2010}
Koprowski, A., Binsztok, H., 2010. {TRX}: A formally verified parser
  interpreter. In: Proceedings of the 19th European Conference on Programming
  Languages and Systems. ESOP'10. Springer-Verlag, Berlin, Heidelberg, pp.
  345--365.
\newline\urlprefix\url{http://dx.doi.org/10.1007/978-3-642-11957-6_19}

\bibitem[{Ramos and de~Queiroz(2015{\natexlab{a}})}]{ramos-2014}
Ramos, M. V.~M., de~Queiroz, R. J. G.~B., Jun. 2015{\natexlab{a}}.
  {Formalization of closure properties for context-free grammars}. ArXiv
  e-prints.
\newline\urlprefix\url{http://arxiv.org/abs/1506.03428}

\bibitem[{Ramos and de~Queiroz(2015{\natexlab{b}})}]{ramos-2015}
Ramos, M. V.~M., de~Queiroz, R. J. G.~B., 2015{\natexlab{b}}. Formalization of
  simplification for context-free grammars. In: Preliminary Proceedings of the
  10th Workshop on Logical and Semantic Frameworks, with Applications. LSFA'15.

\bibitem[{Ramos et~al.(2009)Ramos, Neto, and Vega}]{ramos-2009}
Ramos, M. V.~M., Neto, J.~J., Vega, I.~S., 2009. Linguagens Formais: Teoria
  Modelagem e Implementa\c{c}\~ao. Bookman.

\bibitem[{Ridge(2011)}]{ridge-2011}
Ridge, T., 2011. Simple, functional, sound and complete parsing for all
  context-free grammars. In: CPP. pp. 103--118.

\bibitem[{Sudkamp(2006)}]{sudkamp-2006}
Sudkamp, T.~A., 2006. Languages and {M}achines, 3rd Edition. Addison-Wesley.

\end{thebibliography}
\end {document}